\documentclass[12pt]{article}
\usepackage[cp1251]{inputenc}
\usepackage[russian]{babel}
\usepackage{graphicx}

\def\baselinestretch{1.2}
\begin{document}
\begin{center}
{\Large \bf Critical Behavior of Disordered Systems with a Free Surface} \vspace{0.5cm}

{\bf S. V. Belim\\
Omsk State University, pr. Mira 55, Omsk, 644077 Russia\\
belim@univer.omsk.su}
\end{center}

\vskip 2mm \begin{center} \parbox{142mm} {\small
The behavior of homogeneous and disordered systems with a free boundary is described on the
basis of group theory in the two-loop approximation directly in three-dimensional space. The effect of the free
boundary on the regime of the bulk critical behavior is revealed. It is shown that the boundedness of the system
slightly affects the regime of the bulk critical behavior in the case of the ordinary transition, whereas this effect
is more noticeable in the case of the special transition. Surface critical phenomena are described for homogeneous
and disordered systems, and the critical exponents are calculated in the two-loop approximation. It is
shown that the effect of impurities is insignificant in the special phase transition, whereas it is more noticeable
in the ordinary phase transition. The derived critical exponents are compared with the computer-simulation
results.
}
\end{center} \bigskip

\begin{center}
\end{center}

\section{INTRODUCTION}

Any systems studied experimentally inevitably have free surfaces, whose effect is usually disregarded in
description of critical phenomena. However, ordering processes on the free surface can proceed at a temperature
differing from the temperature characteristic of bulk ordering processes, which leads to change in the
critical behavior regime.

There is a temperature range in which surface effects are decisive and are characterized by a certain
set of the critical exponents.

The problem of the effect of the boundedness of the system on the critical phenomena was first considered
phenomenologically by Kaganov and Omelanchuk \cite{bib1} and in the framework of the microscopic approach by
Mills \cite{bib2} and Wolfram et al. \cite{bib3}. An exact solution for the two-dimensional half-plane was obtained by McCoy
and Wu \cite{bib4} by generalizing Onsager's work \cite{bib5} and showed that the phenomenological theory is valid only
qualitatively. For three-dimensional systems, Binder and Hohenberg \cite{bib6} showed that the separated direction
associated with the existence of the free surface makes necessary a separate description of critical phenomena
on the surface and in the bulk of the system. Binder and Hohenberg \cite{bib6} analyzed two types of the critical behavior
that are possible in semibounded systems and are attributed to the fact that the spin ordering on the surface
occurs earlier than in the bulk (surface transition). In addition, Binder and Hohenberg \cite{bib6} determined the
critical exponents with the use of the renormalization group approach. The critical exponents for semibounded
systems with the scalar order parameter were evaluated in the framework of $\varepsilon$-expansion in the oneloop
approximation by Lubensky and Rubin \cite{bib7}, who obtained the value $\nu=1/2+\varepsilon/12$ for the surface critical
exponent of the correlation radius and the values $\eta_\|=2-\varepsilon/3$ and $\eta_\bot=1-\varepsilon/6$
for the longitudinal and perpendicular critical exponents of the correlation function, respectively. In addition, Lubensky and Rubin \cite{bib8}
showed that all surface critical exponents for semibounded systems with an $n$-dimensional order parameter
are expressed in terms of bulk critical exponents and surface exponent $\widetilde{\eta}=(1/2)\varepsilon (n+2)/(n+8)$
The multicritical point appearing in the simultaneous bulk and phase transitions was analyzed in [9, 10] in the framework
of the \cite{bib9, bib10} $\varepsilon$ expansion in the two-loop approximation.

The surface and bulk phase transitions in a semibounded system with the $n$-dimensional order parameter
were described in \cite{bib11} directly in the three-dimensional space in the two-loop approximation. The critical
exponents calculated in that work are in better agreement with previously obtained Monte Carlo simulation
results \cite{bib17_2, bib17_3, bib17_4}. In particular, for the crossover exponent, computer simulation provides a value of
about $0.5$, calculations directly at $D=3$ yield approximately $0.54$, whereas the $\varepsilon$-expansion gives a result of
about $0.68$.

The renormalization group approach to description of slightly disordered systems developed in \cite{bib15, bib16, bib17}
directly for three-dimensional systems made it possible to obtain the static critical exponents of unbounded systems
in the four-loop approximation. The effect of the free boundary on the critical behavior of disordered systems was first examined
in \cite{bib17_1} in the framework of the $\varepsilon$-expansion. Similar calculations directly in threedimensional
space in the two-loop approximation were performed in \cite{bib17_2, bib17_3, bib17_4}. However, in the latter works, the
replica procedure was performed with an error, owing to which the corresponding asymptotic series for impurity
systems are inconsistent with asymptotic series for homogeneous systems.

In all works mentioned above, it is assumed that the presence of the plane free surface slightly affects the
bulk critical behavior and values obtained for unbounded systems can be used for fixed points of the
renormalization group transformation when calculating the surface critical exponents. However, this statement
should be checked and the effect of the free boundary on the bulk critical behavior should be estimated.

In this work, the effect of the plane free boundary on the bulk critical behavior in various phase transitions in
both homogeneous and disordered systems is examined in the two-loop approximation directly in three-dimensional
space. In addition, the surface critical exponents for the ordinary and special critical exponents for usual
and special phase transitions in homogeneous and disordered systems with the free boundary are calculated
with the inclusion of the corrections obtained.

\section{HAMILTONIAN OF THE SYSTEM}

Let $S=S(\vec{x})$ be fluctuations of the scalar order
parameter in the half-space $V=R^D_+=\{\vec{x}=(\vec{r},z)|\vec{r}\in R^{D-1},z\geq0\}$. The system under consideration is
bounded by the $z=0$ plane, which is denoted as $\partial V$ in what follows. The Hamiltonian of such a model can be written in the form
\begin{eqnarray}\label{gam1}
H_0&&=\frac 12\int_V d^Dx(\tau _0+\nabla^2)S^2(x)
+\frac 12\int_V d^Dx\Delta \tau (x)S^2(x)
+u_0\int_V d^DxS^4(x) \\
&&+\frac {c_0}2\int_{\partial V} d^Dx S^2(x),\nonumber
\end{eqnarray}
Here, $u_0$ is a positive constant; $\tau_0\sim|T-T_c|/T_c$, where $T_c$ is the temperature of the bulk phase transition;
$c_0\sim|T-T_s|/T_s$, where $T_s$ is the temperature of the surface phase transition; and $\Delta\tau(x)$
is the random field of impurities such as random temperature.

Let us pass to Fourier transforms in the $\vec{r}$ coordinates,
\begin{eqnarray}\label{gam2}
&&H_0=\int_0^\infty dz\Big\{\frac 12\int d^{D-1}q(\tau _0+q^2)S_qS_{-q}
+\frac 12\int d^{D-1}q\Delta \tau _qS_qS_{-q} \\
&&+u_0\int d^{D-1}q_1d^{D-1}q_2d^{D-1}q_3S_{q1}S_{q2}S_{q3}S_{-q1-q2-q3}\Big\}\nonumber\\
&&+\frac {c_0}2\int d^{D-1}q S_qS_{-q}\Big|_{z=0}.\nonumber
\end{eqnarray}

For low impurity concentrations, the random field distribution can be treated as Gaussian and be specified
by the function
\begin{eqnarray}
&&P[\Delta \tau]=A\exp [ -\frac 1{2\delta_0}\int \Delta \tau_q^2d^{D-1}qdz],
\end{eqnarray}
where $A$ is the normalization constant and $\delta_0$ is a positive constant proportional to the concentration of frozen
defects of the structure.

The application of the replica procedure for averaging over the random fields specified by the frozen
defects of the structure provides the effective Hamiltonian of the system in the form
\begin{eqnarray}
&&H_{R}=\int_0^\infty dz\Big\{
\frac 12\int d^{D-1}q(\tau _0+q^2)\sum\limits_{a=1}^mS^a _qS^a_{-q}\nonumber \\
&&-\frac{\delta_0}2\sum\limits_{a,b=1}^m\int d^{D-1}q_1d^{D-1}q_2d^{D-1}q_3
S^a_{q1}S^a_{q2}S^b_{q3}S^b_{-q1-q2-q3} \nonumber \\
&&+u_0\sum\limits_{a=1}^m\int d^{D-1}q_1d^{D-1}q_2d^{D-1}q_3S^a_{q1}S^a_{q2}S^a_{q3}S^a_{-q1-q2-q3}\Big\}\nonumber \\
&&+\frac {c_0}2\int d^{D-1}q S_qS_{-q}\Big|_{z=0}.\nonumber
\end{eqnarray}
The properties of the initial system can be obtained in the limit $m\rightarrow0$ for the number of replicas.

The presence of the free boundary even in homogeneous systems leads to the new properties of the model \cite{bib11}. The phase diagram
of such systems exhibits the disordered phase (SD/BD), surface-ordered bulk-disordered phase (SO/BD), and surface-ordered
bulk-ordered phase (SO/BO). The lines on the phase diagram separating these phases determine three kinds
of phase transitions. The transitions from SD/BD to SO/BD, from SO/BD to SO/BO, and from SD/BD to SO/BO are called the surface,
extraordinary, and ordinary phase transitions, respectively. The intersection of these three lines of phase transitions forms the multicritical
point at which the phase transition is called special phase transition. The kind of the phase transition is
definitely determined by the relation between the constants $c_0$ and $\tau_0$. In particular, the condition $\tau_0<<c_0^2<<\Lambda$,
where $\Lambda$ is the inverse lattice parameter and $c_0>0$, is satisfied near the ordinary phase transition;
therefore, the relations $\tau_0\rightarrow 0$ and $c_0/\tau_0\rightarrow \infty$ are valid at the transition point.
The conditions near the extraordinary phase transition have the form $\tau_0<<c_0^2<<\Lambda$ and $c_0<0$, and $\tau_0\rightarrow 0$
and $c_0/\tau_0\rightarrow -\infty$ along the phase transition line. Near the surface phase transition, $c_0<<\tau_0<<\Lambda$, $\tau_0>0$;
therefore, $c_0\rightarrow 0$ and $c_0/\tau_0\rightarrow \infty$ at the phase transition point. For the special phase transition,
$c_0\rightarrow 0$ and $\tau_0\rightarrow 0$; however, $c_0/\tau_0\rightarrow 0$.

As shown in \cite{bib7}, the free propagator in the case under consideration has the form
\begin{equation}\label{kor}
    G(\vec{q};z,z')=\frac1{2k_0}\Big[e^{-k_0|z-z'|}
    -\frac{c_0-k_0}{c_0+k_0}e^{-k_0(z+z')}\Big],
\end{equation}
where $k_0=\sqrt{q^2+\tau_0}$.

\section{BULK CRITICAL PHENOMENA}

The behavior of the system in the critical region is determined by the effective charges at the stationary
point of the renormalization group transformation, which has the following form in the case under consideration:
\begin{eqnarray}
&&S_q^{(0)}=Z^{1/2}S_q,\ \ \
\tau_0=b^2\tau Z_{\tau},\ \ \
u_0=b^{4-D}uZ_u,\ \ \
\delta _0=b^{4-D}\delta Z_\delta,\nonumber
\end{eqnarray}
where $b$ is introduced to reduce the quantities to the dimensionless form.

The $Z$ factors are determined from the requirement of the regularity of the renormalized vertex functions,
which is expressed in the normalization conditions
\begin{eqnarray}
&&Z\frac{\partial}{\partial k^{2}}\Gamma_{bulk}^{(2)}(k) |_{k^{2}=0}=1,\ \ \
Z^{2}{\Gamma_{u,bulk}}^{(4)}|_{k^{2}=0}=b^{4-D}u,\\
&&Z^{2}{\Gamma_{\delta,bulk}}^{(4)} |_{k^{2}=0}=b^{4-D}\delta,\ \ \
Z{\Gamma_{t,bulk}}^{(2,1)}|_{k^{a}=0}=b^{2-D/2}\tau.\nonumber
\end{eqnarray}
Here $\Gamma_{u,bulk}^{(4)}$ and $\Gamma_{\delta,bulk}^{(4)}$ are the four-point bulk vertex functions,
$\Gamma_{bulk}^{(2)}$ is the two-point bulk vertex function, $\Gamma_{t,bulk}^{(2,1)}$ is the two-point bulk vertex function with
the insert.

Let us represent the free propagator in the form of the sum of two terms $G=G_{b}+G_{s}$ . The first term
\begin{equation}
 G_b(\vec{q};z,z')=\frac1{2k_0}e^{-k_0|z-z'|}
\end{equation}
depends only on the difference between the coordinates of two points of the system and, after the Fourier transform
in the $z$ coordinate, acquires the form
\begin{equation}
  G_{b}=\frac 1{\tau_0+\vec{p}^2},
\end{equation}
where $\vec{p}=(\vec{q},p_z)$ is the three-dimensional vector. This expression coincides with the ordinary propagator for
unbounded systems; for this reason, it will be called the bulk propagator in what follows.

The second term
\begin{equation}
 G_s(\vec{q};z,z')=-\frac1{2k_0}\frac{c_0-k_0}{c_0+k_0}e^{-k_0(z+z')}
\end{equation}
depends strongly on the position of both points and is determined by the effective surface charge $c_0$; this part
of the propagator is called the surface propagator. The limiting values of the surface part of the propagator are
represented in the form
\begin{equation}
 G_s^{ord}(\vec{q};z,z')=-\frac1{2k_0}e^{-k_0(z+z')}
\end{equation}
for the case of the ordinary phase transition for $c_0/\tau_0\rightarrow \infty$ and
\begin{equation}
 G_s^{spec}(\vec{q};z,z')=\frac1{2k_0}e^{-k_0(z+z')}.
\end{equation}
for the case of the special phase transition for $c_0/\tau_0\rightarrow 0$.

Since the system is bounded only from one side, the regime of the critical behavior is determined by the
basic volume of the medium that is far from the free surface \cite{bib6}. When considering the points far from the
free surface of the system, one can pass to the limit $z,z'\rightarrow \infty$. In the limiting case under consideration,
$G_b>>G_s$. For this reason, in all previous works devoted to the critical behavior of semibounded systems, small
corrections to the bulk critical behavior that appear due to the presence of the free surface are disregarded. Nevertheless,
the determination of these corrections is an important problem, because any systems studied experimentally are bounded.

Let us write the Callan–Symanzik equation for the bulk vertex functions:
\begin{equation}
[b\frac \partial {\partial b }+\beta _u\frac \partial {\partial
u}+ \beta _\delta \frac \partial {\partial\delta }
-\gamma _\varphi \frac m2b \frac{\partial \ln
Z_\varphi }{\partial b }
-\gamma _\tau \tau \frac \partial {\partial \tau }]\cdot
\Gamma^{(m)}_{bulk}(q;\tau,u ,\delta ,b)=0.
\end{equation}
Here, the functions
\begin{equation}
  \beta_u=b\frac{\partial u}{\partial b},\ \ \ \beta_\delta=b\frac{\partial \delta}{\partial b},\ \ \
  \gamma_\tau=b\frac{\partial \tau}{\partial b},\ \ \ \gamma_\varphi=b\frac{\partial S_q}{\partial b}
\end{equation}
determine the behavior of the system in the critical region.

Figure 1 shows the Feynman diagrams only for the vertex functions of the homogeneous system, which are
necessary for calculation of the corresponding integrals. In these diagrams, the lines corresponding to the
bulk and surface parts of the propagator are marked by the letters "$b$"  and "$s$", respectively.

\begin{figure}
\special{center JETP2005_FIG1.bmp, 16cm 11cm 16cm 11cm} \vspace{12cm}
\caption{Feynman diagrams for the vertex function $\Gamma^{(4)}_u$}\label{fig14}
\end{figure}

The diagrams $J_0^{(0)}$, $J_1^{(0)}$ and $G_0^{(0)}$ contain only "bulk" lines; for this reason, they coincide with the corresponding
integrals for unbounded systems:
\begin{equation}
J_0^{(0)}=\pi^2,\ \ \ J_1^{(0)}=\frac{2\pi^4}{3},\ \ \ G_0^{(0)}=\frac{2\pi^4}{27}.
\end{equation}
The absolute values are given for all other integrals, because their values for the ordinary and extraordinary
transitions differ only in sign:
\begin{eqnarray}
J_0^{(1)}&=&\frac1{(2\pi)^{D-1}}\int_0^\infty\int_0^\infty\int d^{D-1}qdz_1dz_2 G_b(q;z_1,z_2)G_s(q;z_1,z_2)=\frac{5\pi}{16},\\
J_0^{(2)}&=&\frac1{(2\pi)^{D-1}}\int_0^\infty\int_0^\infty\int d^{D-1}qdz_1dz_2 G_s(q;z_1,z_2)G_s(q;z_1,z_2)=\frac{\pi}{16},\nonumber\\
J_1^{(1)}&=&\frac1{(2\pi)^{D-1}}\int_0^\infty\int_0^\infty\int_0^\infty\int \int d^{D-1}qd^{D-1}pdz_1dz_2dz_3 G_b(q;z_1,z_2)\nonumber\\
&&\cdot G_s(-q;z_1,z_3)G_b(p;z_2,z_3)G_b(q-p;z_2,z_3)=1.421210,\nonumber\\
J_1^{(2)}&=&\frac1{(2\pi)^{2D-2}}\int_0^\infty\int_0^\infty\int_0^\infty\int \int d^{D-1}qd^{D-1}pdz_1dz_2dz_3 G_b(q;z_1,z_2)\nonumber\\
&&\cdot G_b(-q;z_1,z_3)G_b(p;z_2,z_3)G_s(q-p;z_2,z_3)=0.760073,\nonumber\\
J_1^{(3)}&=&\frac1{(2\pi)^{2D-2}}\int_0^\infty\int_0^\infty\int_0^\infty\int \int d^{D-1}qd^{D-1}pdz_1dz_2dz_3 G_b(q;z_1,z_2)\nonumber\\
&&\cdot G_b(-q;z_1,z_3)G_s(p;z_2,z_3)G_s(q-p;z_2,z_3)=0.295748,\nonumber\\
J_1^{(4)}&=&\int_0^\infty\int_0^\infty\int_0^\infty\int \int d^{D-1}qd^{D-1}pdz_1dz_2dz_3 G_b(q;z_1,z_2)G_s(-q;z_1,z_3)\nonumber\\
&&\cdot G_b(p;z_2,z_3)G_s(q-p;z_2,z_3)=0.349384,\nonumber\\
J_1^{(5)}&=&\frac1{(2\pi)^{2D-2}}\int_0^\infty\int_0^\infty\int_0^\infty\int \int d^{D-1}qd^{D-1}pdz_1dz_2dz_3 G_b(q;z_1,z_2)\nonumber\\
&&\cdot G_s(-q;z_1,z_3)G_s(p;z_2,z_3)G_s(q-p;z_2,z_3)=0.174535,\nonumber\\
J_1^{(6)}&=&\frac1{(2\pi)^{2D-2}}\int_0^\infty\int_0^\infty\int_0^\infty\int \int d^{D-1}qd^{D-1}pdz_1dz_2dz_3 G_s(q;z_1,z_2)\nonumber\\
&&\cdot G_s(-q;z_1,z_3)G_s(p;z_2,z_3)G_b(q-p;z_2,z_3)=0.177920,\nonumber\\
J_1^{(7)}&=&\frac1{(2\pi)^{2D-2}}\int_0^\infty\int_0^\infty\int_0^\infty\int \int d^{D-1}qd^{D-1}pdz_1dz_2dz_3 G_s(q;z_1,z_2)\nonumber\\
&&\cdot G_s(-q;z_1,z_3)G_s(p;z_2,z_3)G_s(q-p;z_2,z_3)=0.112210,\nonumber\\
G_0^{(1)}&=&-\frac1{(2\pi)^{2D-2}}\frac{\partial}{\partial k^2}\int_0^\infty\int_0^\infty\int \int d^{D-1}qd^{D-1}pdz_1dz_2dz_3 G_b(q+k;z_1,z_2)\nonumber\\
&&\cdot G_b(p;z_1,z_2)G_s(p+q;z_1,z_2)\Big|_{k^2=0}=0.333481,\nonumber\\
G_0^{(2)}&=&-\frac1{(2\pi)^{2D-2}}\frac{\partial}{\partial k^2}\int_0^\infty\int_0^\infty\int \int d^{D-1}qd^{D-1}pdz_1dz_2dz_3 G_b(q+k;z_1,z_2)\nonumber\\
&&\cdot G_s(p;z_1,z_2)G_s(p+q;z_1,z_2)\Big|_{k^2=0}=0.160209,\nonumber\\
G_0^{(3)}&=&-\frac1{(2\pi)^{2D-2}}\frac{\partial}{\partial k^2}\int_0^\infty\int_0^\infty\int \int d^{D-1}qd^{D-1}pdz_1dz_2dz_3 G_s(q+k;z_1,z_2)\nonumber\\
&&\cdot G_s(p;z_1,z_2)G_s(p+q;z_1,z_2)\Big|_{k^2=0}=0.822467,\nonumber\\
J_0^{ord}&=&J_0^{(0)}-2J_0^{(1)}+J_0^{(2)},\nonumber\\
J_0^{spec}&=&J_0^{(0)}+2J_0^{(1)}+J_0^{(2)},\nonumber\\
G_0^{ord}&=&G_0^{(0)}-3G_0^{(1)}+3G_0^{(2)}-G_0^{(3)},\nonumber\\
G_0^{spec}&=&G_0^{(0)}+3G_0^{(1)}+3G_0^{(2)}+G_0^{(3)},\nonumber\\
J_1^{ord}&=&J_1^{(0)}-2J_1^{(1)}-2J_1^{(2)}+J_1^{(3)}+4J_1^{(4)}-2J_1^{(5)}-2J_1^{(6)}+J_1^{(7)},\nonumber\\
J_1^{spec}&=&J_1^{(0)}+2J_1^{(1)}+2J_1^{(2)}+J_1^{(3)}+4J_1^{(4)}+2J_1^{(5)}+2J_1^{(6)}+J_1^{(7)},\nonumber\\
&&J_1^{ord}/(J_0^{ord})^2=0.676982,\ \ \ \ J_1^{spec}/(J_0^{spec})^2=0.747187,\nonumber\\
&&G_0^{ord}/(J_0^{ord})^2=0.065781,\ \ \ \ G_0^{spec}/(J_0^{spec})^2=0.089463.\nonumber
\end{eqnarray}
Without the surface effects, the ratios of the integrals are
\begin{equation}
J_1/(J_0)^2=2/3,\ \ \ \ G/(J_0)^2=2/27.
\end{equation}
Redefining the effective interaction vertices $v_1=u\cdot J_0$ and $v_2=\delta \cdot J_0$,
one arrives at the following expressions for the $\beta-$ and $\gamma$ functions in the two-loop approximation:
\begin{eqnarray}\label{beta}
    \beta_1&=&-(4-D)v_1\Big[1-36v_1+24v_2+1728\Big(2\widetilde{J_1}-1-\frac29\widetilde{G}\Big)v_1^2\nonumber\\
    &-&2304(2\widetilde{J_1}-1-\frac 16\widetilde{G})v_1v_2+672(2\widetilde{J_1}-1-\frac23\widetilde{G})v_2^2\Big],\nonumber\\
    \beta _2&=&-(4-D)v_2 \Big[1-24v_1+8v_2+576(2\widetilde{J_1}-1-\frac 23\widetilde{G})v_1^2\nonumber\\
     &-&1152(2\widetilde{J_1}-1-\frac13\widetilde{G})v_1v_2+352(2\widetilde{J_1}-1-\frac 1{22}\widetilde{G})v_2^2\Big],\nonumber\\
    \gamma_t&=&(4-D)\Big[-12v_1+4v_2+288\Big(2\widetilde{J_1}-1-\frac13\widetilde{G}\Big)v_1^2\nonumber\\
    &-&192(2\widetilde{J_1}-1-\frac 23\widetilde{G})v_1v_2+32(2\widetilde{J_1}-1-\frac12\widetilde{G})v_2^2\Big], \nonumber\\
    \gamma_\varphi&=&(4-D)64\widetilde{G}(3v_1^2-3v_1v_2+v_2^2).\nonumber\\
    \widetilde{J_1}&=&\frac{J_1}{J_0^2}\ \ \ \
    \widetilde{G}=\frac{G}{J_0^2}.\nonumber
\end{eqnarray}

The expressions obtained for the $\beta$-functions are asymptotic series and summation methods must be
used to extract necessary physical information from these series. In this work, the following Borel–Leroy
transformation generalized to the two-parametric case, which provides adequate results for series appearing in
the theory of critical phenomena \cite{bib21_1}, is used:
\begin{equation}
\begin{array}{rl}
  & f(v_1,v_2)=\sum\limits_{i_1,i_2}c_{i_1,i_2}v_1^{i_1}v_2^{i_2}=\int\limits_{0}^{\infty}e^{-t}t^bF(v_1t,v_2t)dt,  \\
  & F(v_1,v_2)=\sum\limits_{i_1,i_2}\frac{\displaystyle c_{i_1,i_2}}{\displaystyle(i_1+i_2+b)!}v_1^{i_1}v_2^{i_2}.
\end{array}
\end{equation}
For the analytic continuation of the Borel transform of a function, the following series in the auxiliary variable $\theta$ is introduced:
\begin{equation}
   {\tilde{F}}(v_1,v_2,\theta)=\sum\limits_{k=0}^{\infty}\theta^k\sum\limits_{i_1,i_2}\frac{\displaystyle c_{i_1,i_2}}{\displaystyle k!}v_1^{i_1}v_2^{i_2}\delta_{i_1+i_2,k},
\end{equation}
to which the [L/M] Pade approximation is applied at the point $\theta=1$. The [2/1] approximants with variation of the parameter $b$
are used to calculate the $\beta$-functions in the two-loop approximation. As shown in \cite{bib21_1}, such variation of $b$
makes it possible to determine the range of variation of the vertex functions and to estimate the accuracy of the critical exponents obtained.

The critical behavior regime is completely determined by the stable stationary points of the renormalization
group transformation, which can be found from the condition that the $\beta$-functions are equal to zero:
\begin{equation}\label{nep}
    \beta_i(v_1^*,v_2^*)=0 \ \ \ \ (i=1,2).
\end{equation}
The requirement of the stability of a fixed point is reduced to the condition that the eigenvalues $b_i$ of the matrix
\begin{equation}
B_{i,j}=\frac{\partial\beta_i(v_1^*,v_2^*)}{\partial{v_j}}\ \ \ \ (i,j=1,2).
\end{equation}
are positive. It is worth noting that the Pade–Leroy summation procedure is possible not for any $b$ values
and this significantly limits the possibility of applying the method. This limitation is associated with the
appearance of the poles of the approximants near the solutions of the system of Eqs. (\ref{nep}); for this reason, it is impossible
to determine the position of the fixed points. In this work, the parameter $b$ varies from $0$ to a value beginning with which
the determination of the stable fixed point becomes impossible. In this range, $20$ values of the parameter $b$ are taken for which the
fixed points are searched. Average values with a certain accuracy determined by the spread in the values for various $b$
values are taken as the effective charges at the fixed point.

The stable stationary point of the renormalization group transformation is determined by the values $(v_1^{ord*}=0.048\pm 0.002$ and $v_2^{ord*}=0$
for the ordinary phase transition in homogeneous systems and by the values $v_1^{spec*}=0.066\pm 0.007$ and $v_2^{spec*}=0$
special phase transition. These quantities for impurity systems are $v_1^{ord*}=0.067\pm 0.002$ and $v_2^{ord*}=0.033\pm 0.003$
for the ordinary phase transition and $v_1^{spec*}=0.071\pm 0.001$ and $v_2^{spec*}=0.015\pm 0.002$ for the special
phase transition. For comparison, the stationary point of the renormalization group transformation in the two-loop approximation is given
by the values $(v_1^{*}=0.046\pm 0.002$ and $v_2^{*}=0)$ for an unbounded homogeneous medium \cite{bib24} and by the values
$v_1^{*}=0.067\pm 0.002$ and $v_2^{*}=0.035\pm 0.003)$ for a medium with frozen structure defects \cite{bib25}.

The exponent $\nu$ characterizing an increase in the correlation radius near the critical point $(R_c\sim|T-T_c|^{-\nu})$
is determined from the relations:
\begin{eqnarray}
  \nu&=&\frac12(1+\gamma_t)^{-1}=\frac12\Big[1+6v_1-v_2-144\Big(2\widetilde{J_1}-1-\frac13\widetilde{G}-\frac14\Big)v_1^2\nonumber\\
    &+&96\Big(2\widetilde{J_1}-1-\frac 12\widetilde{G}-\frac14\Big)v_1v_2-16\Big(2\widetilde{J_1}-1-\frac12\widetilde{G}-\frac14\Big)v_2^2\Big].\nonumber
\end{eqnarray}
The substitution of the effective charges of the fixed points of the special transition yields the values $\nu^{spec}=0.654\pm 0.006$
and $\nu^{spec}_{imp}=0.669\pm 0.003$ for homogeneous systems and impurity systems, respectively. For the ordinary phase transition,
$\nu^{ord}=0.635\pm 0.007$ and $\nu^{ord}_{imp}=0.684\pm 0.002$. For comparison, the critical exponents of the system disregarding the free boundary
for homogeneous systems and impurity systems are $\nu=0.632\pm 0.004$ and $\nu_{imp}=0.685\pm 0.003$, respectively.

The Fisher exponent $\eta$ describing the behavior of the correlation function near a critical point in the space
of wave vectors $(G\sim k^{2+\eta})$ is determined in terms of the scaling function $\gamma_\varphi$ as $\eta=\gamma_\varphi(v_1^*,v_2^*)$.
In the ordinary phase transition in homogeneous systems and impurity systems, $\eta^{ord}=0.029\pm 0.003$ and $\eta^{ord}_{imp}=0.033\pm 0.002$,
respectively; in the special phase transition in homogeneous systems and impurity systems, $\eta^{spec}=0.068\pm 0.008$
and $\eta^{spec}_{imp}=0.069\pm 0.004$, respectively. For comparison, the corresponding Fisher exponents disregarding surface effects
are $\eta=0.030\pm 0.002$ and $\eta_{imp}=0.035\pm 0.001$ for homogeneous systems and impurity systems, respectively. The other critical exponents
can be determined from scaling relations.

\section{SPECIAL PHASE TRANSITION}

In addition to the ordinary critical exponents, which are called bulk exponents in what follows, the system is
characterized by the surface critical exponents on the boundary plane.

Let us introduce the ($N+M$)-point correlation function
\begin{equation}
  G^{(N,M)}(x,r)=\Big<\prod_{i=1}^N S(\vec{x}_i)\prod_{j=1}^M S(\vec{r}_j ,0)\Big>
\end{equation}
with $N$ points under the surface and $M$ points on the medium surface, and $\Big<...\Big>$ means thermodynamic averaging
with the Boltzmann factor $exp(-H[S])$. The transition to the Fourier transforms in the coordinates parallel
to the free boundary yields
\begin{eqnarray}
  &&G^{(N,M)}(\vec{p},z)(2\pi)^{D-1}\delta\Big[\sum_{k=1}^{N+M}\vec{p}_k\Big]=\\
  &&=\int G^{(N,M)}(x,r)\exp\Big[-i\sum_{i=1}^{N}\vec{p}_i\vec{x}_{i||}-i\sum_{j=1}^{M}\vec{p}_{N+j}\vec{r}_{j}\Big]
  \Big[\prod_{i=1}^N d\vec{x}_{i||}\Big]\Big[\prod_{j=1}^M d\vec{r}_{j}\Big]\nonumber
\end{eqnarray}

The renormalization group transformations for quantities on the surface have the form
\begin{eqnarray}
&&S_q^{(s)(0)}=Z_1^{1/2}S^{(s)}_q,\ \ \
c_0=b^2 c Z_{||},
\end{eqnarray}
where $S^{(s)}_q=S(z)_q\Big|_{z=0}$.

To renormalize the correlation function, the following relation can be written:
\begin{equation}
 G^{(N,M)}_{ren}(\tau,u,\delta,c)=Z^{-(N+M)/2}Z_1^{-M/2}G^{(N,M)}(\tau_0,u_0,\delta_0,c_0).
\end{equation}
We introduce the correlation function with inserts in the form
\begin{equation}
 G^{(N,M;I,I_1)}(\tau_0,u_0,\delta_0,c_0)=\frac{\partial^I}{(\partial \tau_0)^I}
 \frac{\partial^{I_1}}{(\partial c_0)^{I_1}}G^{(N,M)}(\tau_0,u_0,\delta_0,c_0).
\end{equation}

The renormalization group transformations for the correlation function with the inserts have the form
\begin{equation}
 G^{(N,M;I,I_1)}_{ren}(\tau,u,\delta,c)=Z^{-(N+M)/2}Z_1^{-M/2}Z_{\tau}^IZ_{||}^{I_1}G^{(N,M;I,I_1)}(\tau_0,u_0,\delta_0,c_0).
\end{equation}
The expansion of $G^{(N,M)}(\tau_0,u_0,\delta_0,c_0)$ in the lowest
expansion order in the powers of $u_0$ has the form
\begin{equation}
 G^{(N,M)}(p;\tau_0,u_0,\delta_0,c_0)=\frac1{c_0+\sqrt{p^2+\tau_0}}+O(u_0,\delta_0).
\end{equation}
Similar to bulk vertex functions, the normalization conditions are taken such that the surface two-point correlation
function for zero external momentum $p=0$ coincides with its zeroth approximation. Correspondingly,
the substitution $\tau_0\rightarrow\tau$, $c_0\rightarrow c$ provides
\begin{eqnarray}
&& G^{(0,2)}_{ren}(p;\tau,u,\delta,c)\Big|_{p=0}=\frac1{\sqrt{\tau}+c},\\
&& \frac{\partial}{\partial p^2}G^{(0,2)}_{ren}(p;\tau,u,\delta,c)\Big|_{p=0}= - \frac1{2\sqrt{\tau}(\sqrt{\tau}+c)^2}.\nonumber
\end{eqnarray}
For the correlation function with the insert, the normalization condition has the form
\begin{equation}
 G^{(0,2;0,1)}(p;\tau,u,\delta,c)\Big|_{p=0}=\frac1{(c+\sqrt{\tau})^2}.
\end{equation}
The surface $Z$ factors $Z_1$ and $Z_{||}$ are determined from the relations
\begin{eqnarray}\label{v2}
&&Z_1Z=-2\sqrt{\tau}(\sqrt{\tau}+c)^2\frac{\partial}{\partial p^2}
G^{(0,2)}(p;\tau_0(\tau,u,\delta),u_0(\tau,u,\delta),\delta_0(\tau,u,\delta),c_0(c,\tau,u))\Big|_{p=0}\\
&&Z_{||}=-Z_1Z(\sqrt{\tau}+c)^2\frac{\partial}{\partial c_0}
G^{(0,2)}(0;\tau_0(\tau,u,\delta),u_0(\tau,u,\delta),\delta_0(\tau,u,\delta),c_0)\Big|_{c_0=c_0(c,\tau,u)}.\nonumber
\end{eqnarray}
In order to simplify the calculations, it is convenient to represent these relations in the form
\begin{eqnarray}
&&(Z_1Z)^{-1}=\lim_{p\rightarrow 0}\frac{\sqrt{\tau}}{p}\frac{\partial}{\partial p}
\Big[G^{(0,2)}(p;\tau_0(\tau,u),u_0(\tau,u),c_0(c,\tau,u))\Big]^{-1},\\
&&Z_{||}^{-1}=Z_1Z\frac{\partial}{\partial c_0}
\Big[G^{(0,2)}(0;\tau_0(\tau,u,\delta),u_0(\tau,u,\delta),\delta_0(\tau,u,\delta),c_0)\Big]^{-1}\Big|_{c_0=c_0(c,\tau,u,\delta)}.\nonumber
\end{eqnarray}
Let us use the decomposition of the correlation function into the free part
\begin{equation}
\Big[G^{(0,2)}(p;\tau_0,0,0,c_0)\Big]^{-1}=c_0+k_0
\end{equation}
and fluctuation corrections $\sigma_0(p;\tau_0,u_0,\delta_0,c_0)$, i.e.,
\begin{equation}\label{1}
  \Big[G^{(0,2)}(p;\tau_0,u_0,\delta_0,c_0)\Big]^{-1}=c_0+k_0-\sigma_0(p;\tau_0,u_0,\delta_0,c_0).
\end{equation}
To calculate $\sigma_0(p;u_0,\delta_0)$, the total propagator between two points on the surface is expressed in terms of the
free energy $\Sigma(p;\tau_0,u_0)$ determining the fluctuation corrections to the two-point vertex function $\Gamma^{(2)}_\tau$ as
\begin{eqnarray}\label{2}
  &&G^{(0,2)}(p;\tau_0,u_0,\delta_0,c_0)=_s|G|_s+_s|GTG|_s,\\
  &&T=\Sigma(1-G\Sigma)^{-1}.\nonumber
\end{eqnarray}
The marks $_s|$ and $|_s$ show that the left and right points of the propagator are localized on the surface. The substitution
of Eq. (\ref{1}) into Eqs. (\ref{2}) yields
\begin{equation}
\sigma_0(p;\tau_0,u_0,\delta_0,c_0)=\frac{g_0Tg_0}{1+_s|G|_sg_0Tg_0}=g_0\Sigma g_0+g_0\Sigma G \Sigma g_0-_s|G|_s(g_0\Sigma g_0)^2+O(\Sigma^3),
\end{equation}
where
\begin{equation}
  g_0(p;z)=(c_0+k_0)G(p;z,0)=e^{-k_0z},\ \ \ k_0=\sqrt{\tau_0+p^2}.
\end{equation}
The Feynman diagrams for $G^{(0,2)}$ without impurity vertices in the two-loop approximation are presented in
Fig. 2 and their analytical expression has the form
\begin{equation}
 \Big[G^{(0,2)}(p;\tau_0,0,0,c_0)\Big]^{-1}=c_0+k_0-\sum_{i=1}^4 C_i(p)-\frac{C_1^2(p)}{c_0+k_0}.
\end{equation}
In this expression, $C_i(p)$ is the contribution from the diagram $B_i$ presented in Fig. 2. In this figure, the line
marked by $D$ means the propagator
\begin{equation}
  G_D(\vec{p},z,z')=\frac1{k_0}\Big[e^{-k_0|z-z'|}-e^{-k_0(z+z')}\Big],\ \ \ \ k_0=\sqrt{1+\vec{p}^2}.
\end{equation}

\begin{figure}
\special{center JETP2005_FIG2.bmp, 16cm 7cm 16cm 7cm} \vspace{8cm}
\caption{Feynman diagrams for the correlation function $G^{(0,2)}$ of the homogeneous system.}
\end{figure}

The surface vertex function is represented in the form $c_0=c+\delta c$, where .c is the shift containing ultraviolet
divergences appearing in analysis of surface quantities similar to the free energy for bulk quantities.
The substitution of Eq. (\ref{1}) into (\ref{2}) yields
\begin{equation}\label{v2}
  Z_1Z\Big[k_0+c+\delta c-\sigma(p)\Big]\Big|_{p=0}=\sqrt{\tau}+c.
\end{equation}
From this relation,
\begin{equation}\label{v3}
  \delta c=\Big\{[Z_1Z]^{-1}-1\Big\}(\sqrt{\tau}+c)+\sigma(0).
\end{equation}
Correspondingly, the correlation function has the form
\begin{equation}\label{v4}
  G^{(0,2)}=\Big[k_0-(\sqrt{\tau}+[Z_1Z]^{-1}(\sqrt{\tau}+c)-(\sigma(p,c+\delta c)-\sigma(0,c+\delta c))\Big]^{-1}.
\end{equation}
This expression provides a scheme for calculating the value of each diagram at $p=0$. This statement is also valid for subdiagrams.

For the special phase transition, setting $c=0$ in Eq. (39) and expanding $\delta c$ in the powers of the effective
charges, one can obtain the first-order correction
\begin{equation}\label{v5}
  \delta c^{(1)}=\Big\{[Z_1Z]^{-1}\Big\}^{(1)}\sqrt{\tau}+\sigma(0).
\end{equation}
The correction obtained is used to determine the first expansion terms for $\sigma(p,c+\delta c)$:
\begin{equation}\label{v6}
  \sigma(p,c+\delta c)\Big|_{c=0}=\sigma(p,0)+\frac{\partial\sigma(p,c)}{\partial c}\Big|_{c=0}\delta\sigma^{(1)}.
\end{equation}

The Callan–Symanzik differential equation of the renormalization group in the critical region has the form \cite{bib11}:
\begin{equation}
\Big[ b\frac{\partial}{\partial b}+\beta_1\frac{\partial}{\partial v_1}+\beta_2\frac{\partial}{\partial v_2}
+\frac{N+M}{2}\eta+\frac{M}{2}\eta_1^{sp}-(1+\eta_c^{sp})\widetilde{c}\frac{\partial}{\partial \widetilde{c}}\Big]G^{(N,M)}_{ren,sp}=0,
\end{equation}
where
\begin{equation}
\widetilde{c}=\frac{c}{\sqrt{\tau}},\ \ \ \eta_1=\Big(\beta_u\frac{\partial}{\partial u}
+\beta_\delta\frac{\partial}{\partial \delta}\Big)\ln Z_1, \ \ \eta_{||}=\Big(\beta_u\frac{\partial}{\partial u}
+\beta_\delta\frac{\partial}{\partial \delta}\Big)\ln Z_{||}.
\end{equation}

The expression for $\eta_{||}^{(spec)}=\eta^{(spec)}+\eta_{1}^{(spec)}$ for disordered system has the form
\begin{eqnarray}
  &&\eta_{||}^{(spec)}=-12\overline{B}_0v_1+12\overline{B}_0v_2
  +288(\overline{B}_1+\overline{B}_2+\frac23\overline{B}_3-\frac32\overline{B}_0+\overline{B}_0\overline{\widetilde{B}}_0-\frac12\overline{B}_0^2)v_1^2\nonumber\\
  &&+288(\overline{B}_1+\overline{B}_2+\frac23\overline{B}_3-\frac23\overline{B}_0+\overline{B}_0\overline{\widetilde{B}}_0-\frac12\overline{B}_0^2)v_2^2\\
  &&-576(\overline{B}_1+\overline{B}_2+\frac23\overline{B}_3+\overline{B}_0\overline{\widetilde{B}}_0-\frac12\overline{B}_0^2)v_1v_2.\nonumber
\end{eqnarray}
Here, as well as in the description of critical phenomena, $v_1$ and $v_2$ are the effective charges and
\begin{eqnarray}
&&\overline{B}_0=\frac{1}{J_0}\frac{\partial}{\partial p^2}B_0\Big|_{p^2=0},\ \ \
\overline{B}_i=\frac{1}{J_0^2}\frac{\partial}{\partial p^2}B_i\Big|_{p^2=0},\\
&&\overline{\widetilde{B_0}}=\frac{1}{J_0}\frac{\partial^2}{\partial p^2 \partial c_0}B_0\Big|_{c_0=0,p=0},
\ \ \ \ \ (i=1,2,3).\nonumber
\end{eqnarray}
The integrals are equal to
\begin{equation}
\overline{B}_0=0.5,\ \ \ \overline{B}_1+\overline{B}_2=0.323414,\ \ \ \overline{B}_3=0.817579, \ \ \
\overline{\widetilde{B_0}}=-0.316936.
\end{equation}

The summation of the asymptotic expressions by the Borel–Leroy method and the substitution of the effective-charge values at the fixed points
of the renormalization group transformation provide the surface critical exponents $\eta_{||}^{(spec)}=-0.239\pm 0.008$
and $\eta_{||imp}^{(spec)}=-0.279\pm 0.009$ for homogeneous systems and impurity systems, respectively. For comparison, the corresponding
critical exponents disregarding the effect of the free surface on the bulk critical behavior are $\eta_{||}^{(spec)}=-0.189\pm 0.001$
and $\eta_{||imp}^{(spec)}=-0.234\pm 0.005$ for homogeneous systems and impurity systems, respectively.

The other critical exponents can be found from the scaling relations \cite{bib11}:
\begin{eqnarray}
&&\eta_{\bot}=0.5(\eta+\eta_{||}),\ \ \ \beta_1=0.5\nu(D-2+\eta_{||}), \gamma_{11}=\nu(1-\eta_{||}), \\
&&\gamma_1=\nu(2-\eta_{\bot}), \ \ \ \delta_1=\frac{D+2-\eta}{D-2+\eta_{||}},\ \ \ \delta_{11}=\frac{D-\eta_{||}}{D-2+\eta_{||}}.\nonumber
\end{eqnarray}
For homogeneous systems, the critical exponents are
\begin{eqnarray}
&&\eta_{\bot}^{spec}=-0.09\pm 0.01,\ \ \ \beta_1^{spec}=0.249\pm 0.009, \ \ \gamma_{11}^{spec}=0.81\pm 0.03, \\
&&\gamma_1^{spec}=1.37\pm 0.05, \ \ \ \delta_1^{spec}=6.5\pm 0.8,\ \ \ \delta_{11}^{spec}=4.3\pm 0.1.\nonumber
\end{eqnarray}
For impurity systems, the surface critical exponents acquire the values
\begin{eqnarray}
&&\eta_{\bot imp}^{spec}=-0.105\pm 0.007,\ \ \ \beta_{1 imp}^{spec}=0.241\pm 0.007, \gamma_{11 imp}^{spec}=0.86\pm 0.02, \\
&&\gamma_{1 imp}^{spec}=1.41\pm 0.04, \ \ \ \delta_{1 imp}^{spec}=6.8\pm 0.4,\ \ \ \delta_{11 imp}^{spec}=4.5\pm 0.1.\nonumber
\end{eqnarray}

\section{ORDINARY PHASE TRANSITION}

The ordinary phase transition requires consideration of the properties of the system in the limit $\widetilde{c}=c/\sqrt{\tau}\rightarrow \infty$.
However, it is very difficult to examine the function $G^{(N,M)}$ in this limit. As shown in \cite{bib20, bib21, bib22, bib23},
renormalization group analysis is significantly simplified for the function
\begin{equation}
 \widehat{G}^{(N,M)}(x,r)=\Big<\prod_{i=1}^N S(\vec{x}_i)\prod_{j=1}^M \partial_n S(\vec{r}_j ,0)\Big>,
\end{equation}
where $\partial_n$ is the normal derivative to the boundary plane.

Let us denote the correlation function as $\widehat{G}^{(N,M)}_\infty=\lim_{\widetilde{c}\rightarrow\infty}\widehat{G}^{(N,M)}$,
fluctuation-correction function $\sigma^{(D)}(p;\tau,u_0,\delta_0)=\sigma(p;\tau,u_o,\delta_0,\infty)$, and the free propagator
\begin{equation}
 G_D=\lim_{\widetilde{c}\rightarrow\infty}G_0=\frac1{2k_0}\Big[e^{-k_0|z-z'|}-e^{-k_0(z+z')}\Big].
\end{equation}
Similar to the case of the special phase transition,
\begin{equation}
 \widehat{G}^{(0,2)}_\infty(p;\tau_0,u_0,\delta_0)=-k+\sigma^{(D)}(p;\tau,u,\delta),
\end{equation}
Here, $\sigma^{(D)}(p;\tau,u_0,\delta_0)=gT(G_D)g$, where $T(G)$ is defined in Eqs. (\ref{2}), and
\begin{equation}
  g(p;z')=e^{-kz'}=\frac{\partial}{\partial z}G_D(p;z,z')\Big|_{z=0}.
\end{equation}
The surface renormalizing factor $Z_{1,\infty}(u,\delta)$, which describes the renormalization group transformation of
the derivative of fluctuations of the order parameter with respect to the normal to the free surface, is introduced as
\begin{equation}
  (\partial_n S)_{ren}=[Z_{1,\infty}Z]^{-1/2}\partial_n S.
\end{equation}
In this case, the renormalization of the correlation function has the form
\begin{equation}
\widehat{G}^{(N,M)}_{\infty,ren}(\{\vec{p}\};\{z_j\};\tau,u,\delta)=Z^{-(N+M)/2}Z_{1,\infty}^{-M/2}
\Big[\widehat{G}^{(N,M)}_{\infty}(\{\vec{p}\};\{z_j\})-\delta^{M,2}_{N,0}\widehat{G}^{(0,2)}_{\infty}(0)\Big].
\end{equation}
The normalization conditions are represented as
\begin{equation}
\widehat{G}^{(0,2)}_{\infty,ren}(0;\tau,u,\delta)=0,\ \ \
\frac{\partial}{\partial p^2}\widehat{G}^{(0,2)}_{\infty,ren}(p;\tau,u,\delta)\Big|_{p^2=0}=-\frac1{2\sqrt{\tau}}.
\end{equation}
The surface renormalization factor $Z_{1,\infty}(u,\delta)$ is determined from the relation
\begin{equation}
  Z_{1,\infty}(u,\delta)Z(u,\delta)=-\lim_{p\rightarrow 0}\frac{\sqrt{\tau}}{p}\frac{\partial}{\partial p}
  \Big[\widehat{G}^{(0,2)}(p)-\widehat{G}^{(0,2)}(0)\Big].
\end{equation}
The scaling function is introduced as
\begin{equation}
  \eta_{1,\infty}=\beta_1\frac{\partial \ln Z_{1,\infty}}{\partial v_1}+\beta_2\frac{\partial \ln Z_{1,\infty}}{\partial v_2}.
\end{equation}
The surface critical exponent $\eta_{||}^{ord}$ for the correlation function parallel to the free surface is determined from
the relation \cite{bib22}
\begin{equation}
  \eta_{||}^{ord}=2+\eta_{1,\infty}(v_1^*,v_2^*)+\eta(v_1^*,v_2^*),
\end{equation}
where $(v_1^*,v_2^*)$ is the fixed point of the renormalization group transformation.

In the two-loop approximation, the scaling function has the form:
\begin{eqnarray}
  &&\eta_{1,\infty}=2-12\overline{B}_0v_1-12\overline{B}_0v_2
  -288(\overline{B}_1+\overline{B}_2+\frac23\overline{B}_3+\frac32\overline{B}_0-\frac12\overline{B}_0^2)v_1^2\nonumber\\
  &&-288(\overline{B}_1+\overline{B}_2+\frac23\overline{B}_3+\frac23\overline{B}_0-\frac12\overline{B}_0^2)v_2^2\\
  &&-576(\overline{B}_1+\overline{B}_2+\frac23\overline{B}_3-\frac12\overline{B}_0^2)v_1v_2,\nonumber\\
  &&\overline{B}^{ord}_0=0.410227\ \ \ \overline{B}^{ord}_1+\overline{B}^{ord}_2=0 \ \ \ \overline{B}^{ord}_3=-0.630378.\nonumber
\end{eqnarray}

The substitution of the effective charges at the fixed point yields the critical exponents $\eta_{||}^{ord}=1.51\pm 0.04$
and $\eta_{|| imp}^{ord}=1.64\pm 0.06$ for homogeneous systems and impurity systems, respectively. The other surface critical
exponents have the values
\begin{eqnarray}
&&\eta_{\bot}^{ord}=0.77\pm 0.02,\ \ \ \beta_1^{ord}=0.79\pm 0.02, \gamma_{11}^{ord}=-0.32\pm 0.01, \\
&&\gamma_1^{ord}=0.78\pm 0.03, \ \ \ \delta_1^{ord}=1.98\pm 0.06,\ \ \ \delta_{11}^{ord}=0.60\pm 0.02,\nonumber\\
&&\eta_{\bot imp}^{ord}=0.84\pm 0.03,\ \ \ \beta_{1 imp}^{ord}=0.91\pm 0.04, \gamma_{11 imp}^{ord}=-0.44\pm 0.04, \\
&&\gamma_{1 imp}^{ord}=0.80\pm 0.03, \ \ \ \delta_{1 imp}^{ord}=1.88\pm 0.05,\ \ \ \delta_{11 imp}^{ord}=0.51\pm 0.03.\nonumber
\end{eqnarray}

\section{CONCLUSIONS}

The calculations reported above reveal the effect of the free boundary on the bulk critical behavior of
homogeneous systems and disordered systems. The critical exponents $\nu^{ord}=0.635\pm 0.007$ and $\eta^{ord}=0.029\pm 0.003$
obtained for the ordinary phase transition in homogeneous systems coincide within the errors with the respective critical exponents
$\nu=0.632\pm 0.004$ and $\eta=0.030\pm 0.002$ obtained disregarding the boundary effects. Therefore, the presence of the free plane
boundary can be disregarded when describing bulk critical phenomena near the point of the ordinary phase transition.

In agreement with expectations, in the special phase transition in homogeneous systems, the effect of the free boundary leads to a more
noticeable difference of the bulk critical exponents $\nu^{spec}=0.654\pm 0.006$ and $\eta^{spec}=0.068\pm 0.008$ from the respective critical
exponents for the unbounded systems. However, the difference between indices begins with the second place after the decimal point,
which is within the errors of experimental methods and cannot be detected experimentally.

A similar pattern is observed for disordered systems. The bulk critical exponents of the system with the free boundary coincide with the critical
exponents of unbounded systems in the ordinary phase transition ($\nu^{ord}_{imp}=0.684\pm 0.002$ and $\eta^{ord}_{imp}=0.033\pm 0.002$ for the
ordinary transition cf. $\nu_{imp}=0.685\pm 0.003$ and $\eta_{imp}=0.035\pm 0.001$ for the transition in the unbounded system).
For the special phase transition ($\nu^{spec}_{imp}=0.669\pm 0.003$ and $\eta^{spec}_{imp}=0.069\pm 0.004$), the difference in the
values of the critical exponents is more noticeable; however, it is also within the experimental errors similar to homogeneous systems.

As mentioned above, the difference between the bulk critical exponents is small so that it can hardly be determined experimentally.
For this reason, the effect of the free boundary can be disregarded when calculating the bulk critical exponents. However, the effect of
the position of the fixed points on the surface critical exponents is much stronger and it is necessary to take into account the shifts of these points.

Table 1 presents the surface critical exponents obtained in this work, as well as the surface critical exponents obtained by disregarding
the shifts of the fixed points in \cite{bib11} and the Monte Carlo computer simulation results for the special phase transition. As seen
in the table, the results obtained in this work by taking into account that the fixed points of the renormalization
group transformation are shifted due to the effect of the free boundary differ from similar results obtained under
the assumption that the boundedness of the system does not affect the bulk critical phenomena. However, the
ranges of the critical exponent values strongly overlap if the errors are taken into account. Comparison with
the Monte Carlo computer simulation shows that the exponent $\beta_1$ obtained in this work is in better agreement
with the computer experiment. The exponent $\gamma_1$ is in good agreement with the computer simulation results
both with and without the effect of the boundary on the bulk critical phenomena.

\begin{table}
\caption{Surface critical exponents for the special transition in homogeneous systems}
\begin{tabular}{|c|c|c|}
\hline
$\beta_1$ & $\gamma_1$&\\
\hline
$0.249\pm 0.09$ & $1.37\pm 0.05$ & This work  \\
$0.257\pm 0.006$& $1.29\pm 0.07$& Diehl \& Shpot\cite{bib11}\\
$0.18(2)$ &$1.41(14)$  & Landau\& Binder \cite{bib26}\\
$0.2375(15)$&$1.328(1)$ & Ruge \& Wagner \cite{bib27}\\
$0.22$ & - & Vandruscolo и др.\cite{bib28}\\
\hline
\end{tabular}
\end{table}

Comparison shows that the surface critical exponents of the special phase transition for homogeneous and disordered systems almost coincide.
Differences that are observed are likely due to a low order of perturbation theory and will decrease with increasing the
number of terms retaining in the expansions of $\beta$- and $\gamma$-functions. Therefore, the effect of point frozen defects
of the structure on surface critical phenomena in the special phase transition is weak and can hardly be detected experimentally.

Table 2 presents the surface critical exponents obtained in this work, as well as the surface critical exponents obtained disregarding
the shift of the fixed points in \cite{bib11} and the Monte Carlo computer simulation results for the ordinary phase transition. According to
the tables, the results obtained in this work by taking into account that the fixed points of the renormalization
group transformation are shifted due to the effect of the free boundary are in good agreement with all results of
the Monte Carlo computer simulation. It is worth noting that the critical exponents of the ordinary surface
transition disregarding the free-surface-induced shift of the fixed points are inconsistent only with the results
reported in [29]. Similar to the special phase transition, the effect of the free surface on the position of the fixed
points cannot likely be detected experimentally. However, in contrast to the special transition, the effect of
impurities in the ordinary phase transition is stronger and can likely be revealed in experiment.

\begin{table}
\caption{Surface critical exponents for the ordinary transition in homogeneous systems}
\begin{tabular}{|c|c|c|}
\hline
$\beta_1$ & $\gamma_1$&\\
\hline
$0.79\pm 0.02$& $0.78\pm 0.03$  & This work  \\
$0.85\pm 0.05$ &$0.74\pm 0.02$& Diehl \& Shpot\cite{bib11}\\
$0.78(2)$ & $0.78(6)$  & Landau\& Binder \cite{bib26}\\
$0.807(4)$& $0.760(4)$ & Ruge \& Wagner \cite{bib27}\\
$0.79(2)$ & - & Kikuchi \& Okabe\cite{bib29}\\
$0.80\pm 0.01$ & $0.78\pm 0.05$ & Pleimling \& Selke\cite{bib30}\\
\hline
\end{tabular}
\end{table}

\section{ACKNOWLEDGMENTS}

This work was supported by the Russian Foundation for Basic Research (project no. 06-02-16018).

\def\baselinestretch{1.0}

\end{document}